\documentstyle[10pt,epsfig,amstex,euscript]{article}
\chardef\anciennecat=\catcode`\@
\catcode`\@=11
\renewcommand\section{\@startsection {section}{1}{\z@}%
                                   {-3.5ex \@plus -1ex \@minus -.2ex}%
                                   {2.3ex \@plus.2ex}%
                                   {\reset@font\large\bfseries}}
\catcode`\@=\anciennecat

\DeclareSymbolFont{AMSb}{U}{msb}{m}{n}
\DeclareMathSymbol{\C}{\mathalpha}{AMSb}{"43}
\DeclareMathSymbol{\R}{\mathalpha}{AMSb}{"52}
\DeclareMathSymbol{\Pom}{\mathalpha}{AMSb}{"50}
\DeclareMathSymbol{\T}{\mathalpha}{AMSb}{"54}

\newcommand{\dpom}{\Delta_{\Pom}}
\newcommand{\as}{\alpha_s}
\newcommand{\ab}{\bar{\alpha}}

\newcommand{\G}{\Gamma}

\newcommand{\ah}{\{\mbox{a.h.}\}}

\newcommand{\real}{{\mathcal R}{\mathrm e}}

\begin{document}

\title{\bf The (BFKL) Pomeron-$\gamma^*$-$\gamma$ vertex for any conformal spin}
\author{S. Munier, H. Navelet\\
 {\it Service de Physique Th\'eorique}\\
 {\it CEA/Saclay}, {\it F-91191 Gif-sur-Yvette Cedex}\\
 {\it France}
}
\maketitle

\begin{abstract} 
To study diffractive photon production at HERA, we compute the
projection of the $\gamma^*\gamma$ impact-factor on the BFKL
leading-order eigenfunctions $E^{n,\nu}$ for non-zero transfer.
This calculation supplements former ones performed for $n=0$.
We provide an expression for $n=\pm 2$ and check that all the other
components are zero.
\end{abstract}

\setcounter{equation}{0}
\setcounter{page}{1}

\section{Introduction.}

The BFKL equation 
has been widely used to study the inclusive or semi-inclusive
observables measured at HERA (structure functions, diffractive 
structure functions) in the small-$x$ (large $s$) 
kinematical region \cite{Forshaw:1997dc}. It is an evolution equation for the gluon
density and was written and solved at leading-$\log(1/x)$ level 
(LLx) \cite{Kuraev:1976ge,Kuraev:1977fs,Balitsky:1978ic}
and at next-to-leading accuracy (NLLx) \cite{Ciafaloni:1998gs,Fadin:1998py}.

In the following, we will be concerned with the LLx kernel, which
exhibits an interesting property:
in the space of the transverse positions $\rho_1$ and $\rho_2$
of the evolved gluons, it was 
shown to be invariant under the global conformal
transformations \cite{Lipatov:1986uk}. Hence the general solution of the
BFKL equation can be written as a sum over the kernel eigenfunctions
corresponding to the irreducible representations of the
symmetry group $SL(2,\C)$. 
The latter are indexed by two indices, one of them being 
the (discrete) conformal spin $n$, the other one the (continuous) real
parameter $\nu$. The dominant energy behaviour is a
power-like rise of the amplitude $s^{\dpom}$, where the intercept
value $\dpom\approx 0.3-0.5$ usually quoted is given by the $n=0$ component.

However, the phenomenological relevance of the higher-spin components ($n\neq 0$)
was suggested in ref.\cite{Munier:1998nt}.
In that paper, it is shown that the $n=\pm 2$ components appearing through
the BFKL resummation can mimic the {\em soft pomeron}. Indeed,
this component exhibits effectively the right
energy dependence of the soft pomeron and reproduces well its ``higher-twist''
behaviour at moderate and large $Q^2$ pointed out by Donnachie and Landshoff 
\cite{Donnachie:1998gm}. Nevertheless, a number of points of the
conjecture were left untested. In particular, only a phenomenological
expression for the coupling of the photon impact-factor to the
higher-spin components was used. Moreover, the non-forward behaviour 
was not considered for lack of a precise knowledge of the relevant
coupling at the photon vertex.

Recently, much efforts have been devoted to the study of the photon
impact-factor in the transition $\gamma^*\rightarrow\gamma$
\cite{Evanson:1999zb,Ivanov:1998jw}. 
Its coupling to the LLx BFKL pomeron was computed but only for $n=0$.
We propose in this paper to extend these calculations to $n=\pm 2$.
The odd components are automatically zero by symmetry. Furthermore,
we have found that the $|n|>2$ ones are zero. This might be due to the
fact that the coupling of two spin-1 photons selects conformal spins
smaller or equal to 2 \cite{robiprive}.
We provide the explicit expressions for $n=0$ and $n=\pm 2$ 
(see eq.(\ref{eq:n2}) and subsequent equations).

The general framework of our calculation is presented in section {\bf
2}. The calculation appears in some details in section {\bf 3}. 
To keep it readable, the most technical point are outlined in the
four appendices {\bf A}-{\bf D} where we show the evaluation of a 
particular generalized hypergeometric ${}_3F_2$ function.
Finally, we draw our conclusions and suggest outlooks.

\section{Definition of the scattering process.}

We consider the scattering of the two objects $1$ and $1^\prime$ into $2$
and $2^\prime$ (see fig.{\bf 1}).
The functions $\phi(\rho_1,\rho_2)$ (resp. $\phi^\prime$) 
are their impact factors and
depend on the bidimensional variables $\rho_1,\rho_2$ which are conjugate to the 
transverse momenta ${\bf k},\ \bf{k+q}$ of the exchanged gluons. 
We deal with an impact factor involving
a virtual photon in the initial state and a real one in the final state.
If furthermore the object $1^\prime$ is a proton, this process models
diffractive photon production at HERA with a large rapidity gap
generated by the BFKL resummation. If $1^\prime$ is also a photon, we have
access to $\gamma^*\gamma^*$ physics. We should not enter into these
details in this paper, and we will only stick to the
$\gamma^*\rightarrow \gamma$ impact-factor.

In the BFKL framework, the most general expression for the scattering
amplitude of two objects $1$ and $1^\prime$ reads \cite{Lipatov:1986uk}:
\begin{equation}
A(s,t)=\frac{i\,s\,G}{(2\pi)^{10}}\sum_{n=-\infty}^{+\infty}
\int d\nu\frac{\nu^2\!+\!\frac{n^2}{4}}
{\left(\nu^2\!+\!\left(\frac{n\!-\!1}{2}\right)^2\right)
\left(\nu^2\!+\!\left(\frac{n\!+\!1}{2}\right)^2\right)}\;e^{\ab
\log\left(s/s_0\right)\chi_n(\nu)}\;I^{n,\nu}\;\bar{I}^{\prime{n,\nu}}\ ,
\end{equation}
where $G$ is the appropriate colour factor corresponding to the
process under consideration, $\ab\equiv\as\,N_c/\pi$ 
and $\chi_n(\nu)$ is the
well-known eigenvalue of the leading-order BFKL kernel:
\begin{equation}
\chi_n(\nu)=2\,\Psi(1)-2\,\real\,\Psi
\left(\frac{1\!+\!|n|}{2}\!+\!i\nu\right)\ .
\end{equation}
The scale $s_0$ is undetermined at LLx.
The functions $I^{n,\nu}$ and $\bar{I}^{\prime{n,\nu}}$
are the ``vertex functions'', i.e. the impact
factors $\phi$ projected on the corresponding eigenfunctions
$E^{n,\nu}$ of the BFKL kernel:
\begin{equation}
I^{n,\nu}=-\frac{1}{4}\int d\rho_1d\bar\rho_1\int d\rho_2d\bar\rho_2
\phi(\rho_1,\rho_2)E^{n,\nu}(\rho_1,\rho_2)\ ,
\label{eq:I1}
\end{equation}
where $\phi(\rho_1,\rho_2)$
is the impact-factor, and
\begin{equation}
E^{n,\nu}(\rho_1,\rho_2)=(-1)^n
\left(\frac{\rho_1-\rho_2}{\rho_1\rho_2}\right)^a
\left(\frac{\bar{\rho}_1-\bar{\rho}_2}
  {\bar{\rho}_1\bar{\rho}_2}\right)^{\tilde{a}}\ .
\label{eq:Enn}
\end{equation}
The convenient notations $a=\frac{1-n}{2}+i\nu$ and 
$\tilde{a}=\frac{1+n}{2}+i\nu$ have been introduced in the previous equation.
We work with complexified transverse vectors $\rho=\rho_x+i\rho_y$ and
$\bar\rho=\rho_x-i\rho_y$.

This impact-factor $\phi$ was computed in ref.\cite{Evanson:1999zb} by
evaluating the relevant Feynman graphs (one of these graphs is depicted in
fig.{\bf 2}). 
Two cases were distinguished. Either the initial off-shell photon scatters
into a real photon of same helicity, or the helicity undergoes a
flip. It was argued that the longitudinal polarization of the
virtual photon does not contribute at LLx. Let us recast the obtained
expressions in the following form:
\begin{multline}
\phi(\rho_1,\rho_2)=-4\pi^2\alpha_e\as e^2\int_0^1
d\alpha f(\alpha)\int dr_1d\bar{r}_1\int dr_2d\bar{r}_2
e^{i\alpha\mathrm{Re}(\bar q r_1)}
e^{i(1-\alpha)\mathrm{Re}(\bar q r_2)}\frac{(r_1\!-\!r_2)^\delta
(\bar{r}_1\!-\!\bar{r}_2)^{\tilde\delta}}
{|r_1\!-\!r_2|^2}\times\\
\times\hat{Q}K_1(|r_1-r_2|\hat{Q})
\left(\delta^2(r_1\!-\!\rho_1)\!-\!\delta^2(r_2\!-\!\rho_1)\right)
\left(\delta^2(r_1\!-\!\rho_2)\!-\!\delta^2(r_2\!-\!\rho_2)\right)\ ,
\label{eq:impact}
\end{multline}
where $\hat{Q}=\sqrt{\alpha(1\!-\!\alpha)}\,Q$, $e^2=\sum_q e_q^2$ 
and $\delta=(1\!-\!\Delta)/2$, $\tilde\delta=(1\!+\!\Delta)/2$.
The expressions for the function $f(\alpha)$ and the exponent $\Delta$ 
depend on the helicity. For the helicity-conserving
processes $(+\rightarrow+)$ and $(-\rightarrow-)$,
$f(\alpha)=\alpha^2\!+\!(1\!-\!\alpha)^2$ and $\Delta=0$. 
In the helicity-flip
case, $f(\alpha)=2\alpha(1\!-\!\alpha)$, $\Delta=-2$ for the 
$(+\rightarrow-)$
transition and $\Delta=+2$ for $(-\rightarrow+)$. The ``$+$'' and
``$-$'' refer to the helicity of the initial state (resp. final state)
photon with respect to the standard basis:
\begin{equation}
\epsilon_\pm=\mp\frac{1}{\sqrt{2}}(1,\pm i)\ .
\end{equation}
The computation of $I^{n,\nu}$ is done in the next section.

\section{Projection on the conformal eigenfunctions.}

Let us now compute the vertex function $I^{n,\nu}$.
Along the lines of ref.\cite{Evanson:1999zb}, one inserts 
eq.(\ref{eq:impact}) into eq.(\ref{eq:I1}).
The product of $\delta$-functions present in the impact-factor
(\ref{eq:impact}) can be expanded. Then two of the terms correspond to the
coupling of the BFKL pomeron to a
single quark line and vanish when projected on the $E^{n,\nu}$.
Indeed, terms of the form $\delta^2(r_1-\rho_1)\delta^2(r_2-\rho_2)$ can be
rearranged to read $\delta^2(\rho_2-\rho_1)\delta^2(2r_1-\rho_1-\rho_2)$, and
they give a zero contribution since 
$E^{n,\nu}(\rho_1,\rho_2)=0$ for $\rho_1=\rho_2$.
Taking into account the symmetry $\alpha\rightarrow 1\!-\!\alpha$ of
$f(\alpha)$, the two remaining contributions are identical and one obtains:
\begin{multline}
I^{n,\nu}=8\pi^2\alpha_e\as e^2\int_0^1d\alpha f(\alpha)\hat{Q}
\int d\rho_1d\bar\rho_1\int d\rho_2d\bar\rho_2E^{n,\nu}(\rho_1,\rho_2)
K_1(|\rho_1-\rho_2|\hat{Q})\times\\
\times e^{i\alpha\real(\bar q\rho_1)}
e^{i(1-\alpha)\real(\bar q\rho_2)}
\frac{(\rho_1\!-\!\rho_2)^\delta
(\bar\rho_1\!-\!\bar\rho_2)^{\tilde\delta}}{|\rho_1\!-\!\rho_2|^2}\ .
\label{eq:I2}
\end{multline}

In the following, the calculation will only be done for positive $n$.
One inserts the expression for $E^{n,\nu}$ in eq.(\ref{eq:I2}) and one 
takes the Mellin representation of the Bessel function:
\begin{equation}
K_1({\mathcal A})=\frac{1}{2}\int\frac{ds}{2i\pi}
\left(\frac{{\mathcal A}}{2}\right)^{-2s-1}
\G(s)\G(1+s)\ ,
\end{equation}
where the contour of integration is parallel to the $z$-axis and
$\real(s)\!>\!0$.
The changes of variable $\rho_1=b(1+t)$ and $\rho_2=b(1-t)$ enable to
reduce one of the \{holomorphic\}$\times$\{antiholomorphic\} integrals in
eq.(\ref{eq:I2}). Indeed, in these new coordinates, it is possible to
factorize and perform the integration over $b$. This integral makes
sense provided that $\gamma\!-\!\tilde\gamma$ is an integer. The
result reads:
\begin{align}
\int db\;d\bar b\,b^{\gamma-1}\bar b^{\tilde\gamma-1}
e^{\frac{i}{2}({\mathcal Q}\bar{b}+\bar{{\mathcal Q}}b)}&=
2i\pi\;e^{i\frac{\pi}{2}(\gamma-\tilde\gamma)}
\frac{\G\left(\frac{\gamma+\tilde\gamma}{2}
\!+\!\frac{|\gamma-\tilde\gamma|}{2}\right)}
{\G\left(1\!-\!\frac{\gamma+\tilde\gamma}{2}
\!+\!\frac{|\gamma-\tilde\gamma|}{2}\right)}
\left(\frac{2}{\bar{{\mathcal Q}}}\right)^\gamma
\left(\frac{2}{{\mathcal Q}}\right)^{\tilde\gamma}\\
&=2i\pi\;e^{i\frac{\pi}{2}(\gamma-\tilde\gamma)}
\frac{\G(\gamma)}{\G(1\!-\!\tilde\gamma)}
\left(\frac{2}{\bar{{\mathcal Q}}}\right)^\gamma
\left(\frac{2}{{\mathcal Q}}\right)^{\tilde\gamma}
\ \ \ \mbox{if $\gamma\!-\!\tilde\gamma\geq 0$}\ ,
\end{align}
with the parameter values ${\mathcal Q}=q(1\!-\!(1\!-\!2\alpha)t)$,
$\gamma=1/2\!-\!a\!-\!s\!+\!\delta$,
$\tilde\gamma=1/2\!-\!\tilde a\!-\!\tilde s\!+\!\tilde\delta$, 
and the convention $\tilde s=s$.
Next, the integral over $t$ can be performed.
It is of the form:
\begin{equation}
\int dt\;
t^{-\frac32+a-s+\delta}(1\!-\!t^2)^{-a}(1\!-\!(1\!-\!2\alpha)t)
^{-\frac12+a+s-\delta} \times\ah\ ,
\end{equation}
where we did not write extensively the antiholomorphic part, 
but it can be obtained by taking the complex
conjugate of $b$ and the ``tilde'' of the exponents.

The conformal mapping $t\rightarrow t/(2-t)$ leads, up to an overall
factor $2^{-a-\tilde a+2s}$, to a well-known holomorphic integral\footnote{
Note that our calculation differs at this level from the ones in 
ref.\cite{Evanson:1999zb,Ivanov:1998jw} but after the integration over
$\alpha$, we obtain the same result, see eq.(\ref{eq:n0}) and appendix
{\bf D} for the comparison.
}.
Here again, to make sense, all the differences $a_i\!-\!\tilde a_i$,
$b_i\!-\!\tilde b_i$ are integer \cite{integrale}:
\begin{multline}
\int dt\;
t^{a_1-1}(1\!-\!t)^{b_1-a_1-1}(1\!-(1-\alpha)t)^{b_0-a_0-1}\times\ah=
2i\frac\mu{\sin\pi b_1}\frac{\pi^2}{\G(b_1\!-\!a_0)\G(b_1\!-\!a_1)}
\Bigg\lbrace
\frac{\G(\tilde a_0)\G(\tilde a_1)}
{\G(\tilde b_1\!-\!\tilde a_0)\G(\tilde b_1\!-\!\tilde a_1)}\times\\
\times\frac{\alpha^{b_1-a_0-a_1}\bar{\alpha}^{\tilde{b}_1-\tilde{a}_0-\tilde{a}_1}}
{\G(1\!-\!a_0)\G(1\!-\!a_1)}
\;{}_2G_1(b_1\!-\!a_0,b_1\!-\!a_1;b_1;1\!-\!\alpha)
{}_2G_1(\tilde b_1\!-\!\tilde a_0,\tilde b_1\!-\!\tilde a_1;\tilde
b_1; 1\!-\!\bar\alpha)
-\frac{(1\!-\!\alpha)^{1-b_1}(1\!-\!\bar{\alpha})^{1-\tilde b_1}}
{\G(1\!-\!b_1\!+\!a_0)\G(1\!-\!b_1\!+\!a_1)}\times\\
\times{}_2G_1(a_0\!-\!b_1\!+\!1,a_1\!-\!b_1\!+\!1;2\!-\!b_1;1\!-\!\alpha)
{}_2G_1(\tilde a_0\!-\!\tilde b_1\!+\!1,\tilde a_1\!-\!\tilde
b_1\!+\!1;2\!-\!\tilde b_1;1\!-\!\bar\alpha)
\Bigg\rbrace\ ,
\label{eq:conforme}
\end{multline}
where
${}_2G_1({\mathcal A},{\mathcal B};{\mathcal C};z)
\equiv\G({\mathcal A})\G({\mathcal B})/\G({\mathcal C})\times
{}_2F_1({\mathcal A},{\mathcal B};{\mathcal C};z)$, and
\begin{equation}
\mu=(-1)^{a_0-\tilde a_0}\frac{\G(\tilde b_1\!-\!\tilde a_1)}
{\G(1\!-\!b_1\!+\!a_1)}\frac{\G(1\!-\!\tilde a_0)}{\G(a_0)}\ .
\end{equation}
In our case, the parameter values are
$a_0=1/2-a-s+\delta$,
$a_1=-1/2+a-s+\delta$,
$b_0=1$,
$b_1=1/2-s+\delta$.
One sees that the convergence of the integral (\ref{eq:conforme})
imposes to chose the integration contour in $s$ such that
$\real(s)<1/2$.

At this stage, a comment on the possible values for $n$ is in order.
Thanks to the relation $a_0\!+\!a_1\!=\!2b_1\!-\!1$, the 
conformal mapping $t\rightarrow -t/(t-1)$ applied to
eq.(\ref{eq:conforme}) leads to the same solution again but for a
factor $(-1)^n$ and the interchange $1\!-\!\alpha\leftrightarrow\alpha$. As
we integrate over $\alpha$ and as all the other factors depending on
$\alpha$ are symmetric under the exchange 
$\alpha\leftrightarrow 1\!-\!\alpha$, we see that the final expression
will have a factor $(1+(-1)^n)/2$, and thus only the even $n$
contribute. This is expected since the BFKL pomeron has only even
conformal spin-components \cite{Navelet:1997tx}. Nevertheless, our
method applies for any $n$ and the final result will be a further check
of the validity of the whole calculation.

All in all, we arrive at the following intermediate expression
(we used $\delta\!+\!\tilde\delta\!=\!1$):
\begin{multline}
I^{n,\nu}=32\pi^3(-1)^{\frac{n}{2}+\frac{\delta-\tilde\delta}{2}+\frac12+\delta}
\alpha_e\as e^2\left(\frac
Q2\right)^{a+\tilde a-2}\sin\pi a\frac{\G(1\!-\!\tilde a)}{\G(a)}
\int\frac{ds}{2i\pi}\frac{\G(s)\G(1\!+\!s)}
{\sin\pi s}\times\\
\times\frac{1}{\G(\frac32\!-\!\delta\!-\!a\!+\!s\!)
\G(\frac12\!-\!\delta\!+\!a\!+\!s)}
\left(\frac{q}{Q}\right)^{-\frac12+\tilde a+s-\tilde\delta}
\left(\frac{\bar{q}}{Q}\right)^{-\frac12+a+s-\delta}
\Bigg\lbrace
\G(-\frac12\!+\!\tilde\delta\!+\!\tilde a\!-\!s)
\G(\frac12\!+\!\tilde\delta\!-\!\tilde a\!-\!s)
\sin\pi\tilde a\times\\
\times\int_0^1d\alpha\;f(\alpha)\alpha^{s}(1\!-\!\alpha)^{-s}\;
{}_2G_1(a,1\!-\!a;\frac12\!-\!s\!+\!\delta;1\!-\!\alpha)
{}_2G_1(\tilde a,1\!-\!\tilde a;\frac12\!-\!s\!+\!\tilde\delta;1\!-\!\alpha)
-\G(\frac32\!-\!\delta\!-\!a\!+\!s\!)
\G(\frac12\!-\!\delta\!+\!a\!+\!s)\times\\
\times\sin\pi a \int_0^1d\alpha\;f(\alpha)\alpha^{-s}(1\!-\!\alpha)^{s}\;
{}_2G_1(a,1\!-\!a;\frac32\!+\!s\!-\!\delta;1\!-\!\alpha)
{}_2G_1(\tilde a,1\!-\!\tilde a;\frac32\!+\!s\!-\!\tilde\delta;1\!-\!\alpha)
\Bigg\rbrace\ .
\end{multline}
The integrations over $\alpha$ remain to be performed. We can treat
all the cases by computing the following generic integral:
\begin{multline}
{\mathcal J}^m(a,c)=\frac{\sin\pi a}{\pi}\frac{\sin\pi \tilde c}{\pi}
\G(\tilde c\!-\!\tilde a)\G(\tilde c\!+\!\tilde a\!-\!1)\times\\
\times\int_0^1d\alpha\;\alpha^{m+1-c}(1-\alpha)^{c-1}
\;{}_2G_1(a,1\!-\!a;c;1\!-\!\alpha)\;
{}_2G_1(\tilde a,1\!-\!\tilde a;\tilde c;1\!-\!\alpha)\ ,
\label{eq:int}
\end{multline}
where $c=1/2\!-\!s\!+\!\delta$ for the first integral and 
$c=3/2\!+\!s\!-\!\delta$ for the second one.
For simplicity, we consider that when one of the arguments of
${\mathcal J}$ has a ``tilde'' it means that
we exchange in the formula the corresponding argument with its ``tilde''
counterpart.
This notation can be slightly misleading since the ``tilde''-operation is not
involutive. 
The index $m\in\{0,1,2\}$ has been introduced in order to take into account
the functions $f(\alpha)$ which we rewrite $1\!-\!2\alpha\!+\!2\alpha^2$ and
$2\alpha(1\!-\!\alpha)$ respectively. Thus the ``physical'' integrals to
compute are:
\begin{align}
{\mathcal J}_{++}(a,c)&={\mathcal J}^0(a,c)-2{\mathcal J}^1(a,c)
+2{\mathcal J}^2(a,c)\nonumber\\
{\mathcal J}_{+-}(a,c)&=2{\mathcal J}^2(a,c)\ .
\end{align}
With these notations, the amplitudes can be written:
\begin{multline}
I^{n,\nu}_{h_1 h_2}=32\pi^4(-1)^{\frac{n}{2}\!+\!
  \frac{\delta\!-\!\tilde\delta}{2}\!+\!\frac12\!+\!\delta}\alpha_e\as 
e^2\left(\frac Q2\right)^{a+\tilde a-2}\sin\pi a\frac{\G(1\!-\!\tilde a)}{\G(a)}
\int\frac{ds}{2i\pi}\frac{\G(s)\G(1\!+\!s)}
{\sin\pi s}\times\\
\times\frac{1}{\G(\frac32\!-\!\delta\!-\!a\!+\!s\!)
\G(\frac12\!-\!\delta\!+\!a\!+\!s)}
\left(\frac{q}{Q}\right)^{-\frac12\!-\!\tilde\delta\!+\!\tilde a\!+\!s}
\left(\frac{\bar{q}}{Q}\right)^{-\frac12\!-\!\delta\!+\!a\!+\!s}
(-1)^n\times\\
\times\Bigg\lbrace
\frac{\pi}{\sin\pi(\frac12\!+\!\tilde\delta\!-\!s)}
{\mathcal J}_{h_1h_2}(a,\frac12\!+\!\delta\!-\!s)-
\frac{\pi}{\sin\pi(\frac32\!-\!\delta\!+\!s)}
{\mathcal J}_{h_1h_2}(1\!-\!\tilde a,
\frac32\!-\!\tilde\delta\!+\!s)
\Bigg\rbrace\ ,
\label{eq:intermediaire}
\end{multline}
where $h_1,h_2 \in \{+,- \}$.

We could directly integrate eq.(\ref{eq:int}) using ref.\cite{prudnikov}.
However, it would lead to ${}_4G_3$ functions tedious to reduce to
elementary functions. Instead, we have a method which leads directly
to the ${}_3G_2$ functions that we explicitly compute in the appendix.
We express ${}_2G_1(z)$ as a
$G^{22}_{22}(1\!-\!z)$ Meijer function, namely:
\begin{align}
{}_2G_1({\mathcal A},{\mathcal B};{\mathcal C};z)&=
\frac{1}{\G({\mathcal C}\!-\!{\mathcal A})
\G({\mathcal C}\!-\!{\mathcal B})}\,
G^{22}_{22}\left(\matrix {1\!-\!{\mathcal A},1\!-\!{\mathcal B}}\\
{\mathcal C\!-\!A\!-\!B}\endmatrix ; 1-z \right)\nonumber\\
&=\frac{1}{\G({\mathcal C}\!-\!{\mathcal A})
\G({\mathcal C}\!-\!{\mathcal B})}
\int\frac{ds^\prime}{2i\pi}\;(1\!-\!z)^{s^\prime}\;\G(-s^\prime)
\G({\mathcal C}\!-\!{\mathcal A}\!-\!{\mathcal B}\!-\!s^\prime)
\G({\mathcal A}\!+\!s^\prime)\G({\mathcal B}\!+\!s^\prime)\ .
\end{align}
Inserting this identity into eq.(\ref{eq:int}) then performing the
integration over $\alpha$ \cite{gradsteyn}, one obtains:
\begin{equation}
{\mathcal J}^m(a,c)=\frac{\sin\pi\tilde c}{\pi}
\int\frac{ds^\prime}{2i\pi}\G(\tilde a\!+\!s^\prime)
  \G(1\!-\!\tilde a\!+\!s^\prime)
\left\{\frac{\G(-s^\prime)\G(m\!+\!1\!+\!s^\prime)
\G(\tilde c\!-\!1\!-\!s^\prime)\G(m\!+\!2\!-\!c\!+\!s^\prime)}
{\G(m\!+\!2\!-\!a\!+\!s^\prime)\G(m\!+\!1\!+\!a\!+\!s^\prime)}
\right\}\ .
\end{equation}
Next, we transform this integral into a sum of ${}_3G_2$ functions,
which is done by reducing the quotient of $\G$-functions between
the brackets in the
r.h.s. and then constructing the defining series for the ${}_3G_2$ by
picking the poles.
This goes as follows. One writes:
\begin{align}
\Bigg\lbrace...\Bigg\rbrace&=\frac{\pi}{\sin\pi(-s^\prime)}\frac{\pi}{\sin\pi(\tilde
c\!-\!1\!-\!s^\prime)}
\left[
\frac{\prod_{j=0}^{m-1}(1\!+\!s^\prime\!+\!j)
\;\times\;{\G(m\!+\!2\!-\!c\!+\!s^\prime)}/
{\G(2\!-\!\tilde c\!+\!s^\prime)}}
{\prod_{j=0}^{k-1}(m\!+\!1\!+\!a\!+\!s^\prime\!-\!k\!+\!j)
(m\!+\!2\!-\!a\!+\!s^\prime\!-\!k\!+\!j)}
\right]\times\nonumber\\
&\hskip 3cm\times\frac{1}{\G(m\!+\!1\!+\!a\!+\!s^\prime\!-\!k)
\G(m\!+\!2\!-\!a\!+\!s^\prime\!-\!k)}\ .
\end{align}
By simple inspection, we get for any $m$:
\begin{equation}
\Bigg\lbrace...\Bigg\rbrace=\frac{\pi}{\sin\pi(-s^\prime)}\frac{\pi}{\sin\pi(\tilde
c\!-\!1\!-\!s^\prime)}
\sum_{p,q=0}^m\frac{{\mathcal
A}_{pq}(a,c)}{\G(1+a+p+s^\prime)\G(2-a+q+s^\prime)}\ ,
\end{equation}
where the ${\mathcal A}_{pq}$s do not depend on $s^\prime$.
The integral defining ${\mathcal J}$ can then be computed by
constructing two series whose coefficients are the residues at the
right poles of the two inverse-sines.
Each of the series is a ${}_3G_2$-function of the type of those computed in
appendix {\bf A}.
The values for the non-vanishing coefficients ${\mathcal A}_{pq}$ we
have chosen are:
\begin{align}
{\mathcal A}_{00}&=1,\ 
{\mathcal A}_{01}=\frac{2}{2a\!-\!1}\left(a(a\!-\!1)\!+\!c(2\!-\!a)\!+\!c^2\right),\ 
{\mathcal A}_{10}=\frac{2}{1\!-\!2a}\left(a(a\!-\!1)\!+\!c(a\!+\!1)\!+\!c^2\right),\ \nonumber\\
{\mathcal A}_{12}&=\frac{2}{2a\!-\!1}(a\!-\!1)(a\!-\!2)(a\!-\!c)(a\!-\!c\!-\!1),\ 
{\mathcal A}_{21}=\frac{2}{1\!-\!2a}a(a\!+\!1)(a\!+\!c)(a\!+\!c\!-\!1)
\end{align}
for the helicity-conserving $(+\rightarrow +)$ amplitude and
\begin{align}
{\mathcal A}_{11}&=2,\ 
{\mathcal A}_{12}=\frac{2}{2a\!-\!1}(a(a\!-\!3)\!+\!2),\ 
{\mathcal A}_{21}=\frac{2}{1\!-\!2a}a(a\!+\!1)
\end{align}
for the helicity-flip $(+\rightarrow -)$ one.
Let us complete the calculation for the ${\mathcal J}$'s.
\begin{equation}
{\mathcal J}_{h_1h_2}(a,c)=\sum_{p,q}{\mathcal A}_{h_1h_2,pq}(a,c)\left\{
{}_{3}G_{2}\!\left(\matrix {1,\tilde c\!+\!\tilde a\!-\!1,\tilde c\!-\!\tilde a}\\
  {\tilde c\!+\!a\!+\!p, \tilde c\!-\!a\!+\!1\!+\!q}\endmatrix ;
  {\displaystyle 1}\right)
-{}_{3}G_{2}\!\left(\matrix {1,\tilde a,1\!-\!\tilde a}\\ 
{1\!+\!a\!+\!p, 2\!-\!a\!+\!q}\endmatrix ;
  {\displaystyle 1}\right)
\right\}\ ,
\label{eq:diffG}
\end{equation}
where we have been able to factorize and simplify $\pi/\sin\pi\tilde
c$ since $c$ and $\tilde c$ differ by an even integer ($0$ or $2$) in all cases.
The differences of ${}_3G_2$ in the r.h.s. can be computed using the tricks exposed
in appendix~{\bf A} and {\bf B}.
The ones needed for our purpose are explicitly shown in
appendix~{\bf C}.

However, formula (\ref{eq:res}) enables us to perform the calculation
for general $n$ (and then take the appropriate limits). 
It enables us to write the ${\mathcal J}$s
as a sum of rational fractions in $a$, $\tilde a$ and
$c$, times a non-rational coefficient which is either $\sin\pi
a/\sin\pi\tilde a$ or ${\G(c\!-\!\tilde a)\G(c\!+\!\tilde a\!-\!1)}/
{\G(c\!-\!a)\G(c\!+\!a\!-\!1)}$.
The rational fractions are tedious but straightforward to compute using
{\tt mathematica}. Inserting eq.(\ref{eq:diffG}) after appropriate
computation into eq.(\ref{eq:intermediaire}),
and recalling that $a=(1\!-\!n)/2\!+\!i\nu$ yield for $n=0$:
\begin{align}
{\mathcal J}_{++}(\frac12\!+\!i\nu,1\!-\!s)+{\mathcal
J}_{++}(\frac12\!-\!i\nu,1\!+\!s)=&
\frac{\pi}{16i\nu(1\!+\!\nu^2)}\frac{\sin^2\pi s\tan\pi i\nu}
{\cos\pi(i\nu\!-\!s)\cos\pi(i\nu\!+\!s)}\times\nonumber\\
 &\ \ \ \times(11\!+\!12\nu^2\!+\!4s^2)\nonumber\\
{\mathcal J}_{+-}(\frac12\!+\!i\nu,2\!-\!s)+{\mathcal
J}_{+-}(\frac12\!-\!i\nu,2\!+\!s)=&
\frac{\pi}{4i\nu(1\!+\!\nu^2)}\frac{\sin^2\pi s\tan\pi i\nu}
{\cos\pi(i\nu\!-\!s)\cos\pi(i\nu\!+\!s)}\ .
\end{align}
Those corresponding to $n=2$ read:
\begin{align}
{\mathcal J}_{++}(-\frac12\!+\!i\nu,1\!-\!s)+{\mathcal
J}_{++}(-\frac12\!-\!i\nu,1\!+\!s)=&
-\frac{\pi}{32i\nu(1\!+\!\nu^2)}
\frac{\sin^2\pi s\tan\pi i\nu}
{\cos\pi(i\nu\!-\!s)\cos\pi(i\nu\!+\!s)}\times\nonumber\\
 &\ \ \ \times(-1\!+\!2i\nu\!-\!2s)(1\!+\!2i\nu\!-\!2s)\nonumber\\
{\mathcal J}_{+-}(-\frac12\!+\!i\nu,2-s)+{\mathcal
J}_{+-}(-\frac12\!-\!i\nu,2\!+\!s)=&
-\frac{\pi}{8i\nu(1\!+\!\nu^2)}
\frac{\sin^2\pi s\tan\pi i\nu}
{\cos\pi(i\nu\!-\!s)\cos\pi(i\nu\!+\!s)}\ .
\end{align}
As expected, we find that all the other components are zero.

Let us summarize our final results.

We obtain the selection rule that for a virtual photon scattering, only
the components $n=0$ and $n=\pm 2$ contribute. We list
underneath the expressions for these non-vanishing amplitudes. We
introduce the angle $\phi$ between the plane formed by the initial- and final-state
electrons and the initial- and final-state photons
respectively, which is the argument of $q$.
First, the $n=0$ components:
\begin{multline*}
I_{++}^{0,\nu}=-2\pi^6\alpha_e\as e^2\left(\frac
Q2\right)^{-1\!+\!2i\nu}\frac{\tanh\pi\nu}{\pi\nu(\nu^2+1)}
\frac{1}{\G^2(\frac12\!+\!i\nu)}
\times\int\frac{ds}{2i\pi}
\left(\frac{q^2}{Q^2}\right)^{-\frac12\!+\!i\nu\!+\!s}\times\\
\times\G(s)\;\G(1\!+\!s)\;
\G(\frac12\!-\!i\nu\!-\!s)\;
\G(\frac12\!+\!i\nu\!-\!s)\;
\times\left(11\!+\!12\nu^2\!+\!4s^2\right)
\end{multline*}

\begin{multline}
I_{+-}^{0,\nu}=8\pi^6\alpha_e\as e^2\left(\frac
Q2\right)^{-1\!+\!2i\nu}\frac{\tanh\pi\nu}{\pi\nu(\nu^2\!+\!1)}
\frac{1}{\G^2(\frac12+i\nu)}e^{2i\phi}
\times\int\frac{ds}{2i\pi}
\left(\frac{q^2}{Q^2}\right)^{-\frac12\!+\!i\nu\!+\!s}\times\\
\times\G(s)\;\G(1\!+\!s)\;
\G(\frac32\!-\!i\nu\!-\!s)\;
\G(\frac32\!+\!i\nu\!-\!s)\ .
\label{eq:n0}
\end{multline}
We checked that these two expressions agree with
ref.\cite{Evanson:1999zb} but for an overall sign difference.

\noindent Second, the $n=2$ components:
\begin{multline*}
I_{++}^{2,\nu}=-4\pi^6\alpha_e\as e^2\left(\frac
Q2\right)^{-1\!+\!2i\nu}\frac{\tanh\pi\nu}{\pi\nu(\nu^2\!+\!1)}
\frac{1}{\G^2(\frac12\!+\!i\nu)}
\frac{-\frac12\!+\!i\nu}{+\frac12\!+\!i\nu}
e^{2i\phi}
\times\int\frac{ds}{2i\pi}
\left(\frac{q^2}{Q^2}\right)^{-\frac12\!+\!i\nu\!+\!s}\times\\
\times\G(s)\;\G(1\!+\!s)\;
\G(\frac32\!-\!i\nu\!-\!s)\;
\G(\frac32\!+\!i\nu\!-\!s)\;
\end{multline*}

\begin{multline}
I_{+-}^{2,\nu}=4\pi^6\alpha_e\as e^2\left(\frac
Q2\right)^{-1\!+\!2i\nu}\frac{\tanh\pi\nu}{\pi\nu(\nu^2\!+\!1)}
\frac{1}{\G^2(\frac12\!+\!i\nu)}
\frac{-\frac12\!+\!i\nu}{+\frac12\!+\!i\nu}
e^{4i\phi}
\times\int\frac{ds}{2i\pi}
\left(\frac{q^2}{Q^2}\right)^{-\frac12\!+\!i\nu\!+\!s}\times\\
\times\G(s)\;\G(1\!+\!s)\;
\G(\frac12\!+\!i\nu\!-\!s)\;
\G(\frac52\!-\!i\nu\!-\!s)\ .
\label{eq:n2}
\end{multline}
Third, the $n=-2$ components are deduced from the preceeding ones
using an appropriate relation between $E^{n,\nu}$ and $E^{-n,\nu}$. 
Let us briefly derive this relation. We come back to
eq.(\ref{eq:I2}) and note that it factorizes in the following 
way by performing the change of variable $\rho_1=b\!+\!\rho/2$ 
and $\rho_2=b\!-\!\rho/2$:
\begin{equation}
I^{n,\nu}=\int_0^1d\alpha\int d\rho\, d\bar\rho\,\, {\mathcal
H}(\alpha,\rho)\, \pi^3\frac{2^{4i\nu}}{-i\nu\!+\!n/2}\,
\frac{\G(-i\nu\!+\!(1\!+\!n)/2)}{\G(i\nu\!+\!(1\!+\!n)/2)}\,
\frac{\G(i\nu\!+\!n/2)}{\G(-i\nu\!+\!n/2)} E^{n,\nu}_q(\rho)\ ,
\end{equation}
where $E^{n,\nu}_q(\rho)$ is defined in
ref.\cite{Lipatov:1986uk}. We put in the function ${\mathcal
H}(\alpha,\rho)$ all the other dependencies, namely:
\begin{equation}
{\mathcal H}(\alpha,\rho)=8i\,\alpha_e\alpha_s\,e^2\,f(\alpha)\hat{Q}\,
e^{i(\alpha-1/2)\real(\bar q\rho)}K_1(|\rho|\hat{Q})
\frac{\rho^\delta{\bar\rho}^{\tilde\delta}}{|\rho|}\ .
\end{equation}
Note that this formula shows that $I^{n,\nu}$ is the projection of
the impact factor on $E_q^{n,\nu}$ \cite{Navelet:1997qz}.
We then notice that for positive $n$ \cite{Lipatov:1986uk}:
\begin{equation}
E_q^{-n,\nu}(b)=2^{-12i\nu}\,\frac{n/2\!-\!i\nu}{n/2\!+\!i\nu}
\,\frac{\G^2(n/2\!-\!i\nu)}{\G^2(n/2\!+\!i\nu)}\,
q^{2a\!-\!1}\bar{q}^{2\tilde a\!-\!1}E_q^{n,-\nu}(b)\ ,
\end{equation}
and hence we arrive at the relation:
\begin{equation}
I^{-n,\nu}=2^{-4i\nu}\,\frac{\G^2(-i\nu\!+\!(1\!+\!n)/2)}
{\G^2(i\nu\!+\!(1\!+\!n)/2)}\,
q^{2a\!-\!1}\bar{q}^{2\tilde a\!-\!1}\,
\;I^{n,-\nu}\ .
\label{eq:relnmn}
\end{equation}
Applying this relation to eq.(\ref{eq:n2}), one obtains:
\begin{multline*}
I_{++}^{-2,\nu}=-4\pi^6\alpha_e\as e^2\left(\frac
Q2\right)^{-1\!+\!2i\nu}\frac{\tanh\pi\nu}{\pi\nu(\nu^2\!+\!1)}
\frac{1}{\G^2(\frac12\!+\!i\nu)}
\frac{-\frac12\!+\!i\nu}{+\frac12\!+\!i\nu}
e^{-2i\phi}
\times\int\frac{ds}{2i\pi}
\left(\frac{q^2}{Q^2}\right)^{-\frac12\!+\!i\nu\!+\!s}\times\\
\times\G(s)\;\G(1\!+\!s)\;
\G(\frac32\!-\!i\nu\!-\!s)\;
\G(\frac32\!+\!i\nu\!-\!s)\;
\end{multline*}

\begin{multline}
I_{+-}^{-2,\nu}=4\pi^6\alpha_e\as e^2\left(\frac
Q2\right)^{-1\!+\!2i\nu}\frac{\tanh\pi\nu}{\pi\nu(\nu^2\!+\!1)}
\frac{1}{\G^2(\frac12\!+\!i\nu)}
\frac{-\frac12\!+\!i\nu}{+\frac12\!+\!i\nu}
\times\int\frac{ds}{2i\pi}
\left(\frac{q^2}{Q^2}\right)^{-\frac12\!+\!i\nu\!+\!s}\times\\
\times\G(s)\;\G(1\!+\!s)\;
\G(\frac12\!-\!i\nu\!-\!s)\;
\G(\frac52\!+\!i\nu\!-\!s)\ .
\end{multline}
Note that the complex integrals over $s$ can be expressed with (one or
a sum of) Legendre functions.
Note also that the helicity-flip component for $n=2$ does not vanish
at small $q$.

The other helicity amplitudes are simply obtained using the relations
$I_{--}^{n,\nu}=I_{++}^{-n,\nu}\vert_{\phi\rightarrow-\phi}$ and
$I_{-+}^{n,\nu}=I_{+-}^{-n,\nu}\vert_{\phi\rightarrow-\phi}$.


\section{Conclusion.}

We have computed the coupling of a $\gamma^*\gamma$ impact-factor to
the LLx BFKL pomeron. We have found that our calculation is consistent
with a previous one \cite{Evanson:1999zb} for the $n=0$ component.
For the higher-conformal spin components, only the two values $n=\pm 2$
contribute, as expected from symmetry considerations. 

The physical motivation underneath was the precise study of diffractive photon
production in the high-energy regime, including the ``higher-twist''
type components induced by conformal invariance \cite{Munier:1998nt}.
With the results we obtained in this paper, we are almost ready to
study the phenomenology of the higher-spin components of the BFKL
pomeron with a realistic impact-factor.
It should also be worth to transpose the methods developped here to the
computation of other impact factors.
We leave these studies for forthcoming papers.
\\


\vskip 0.5cm
\noindent
{\bf Acknowledgements}\\
We thank R.Peschanski for helpful comments and a careful reading of
the manuscript, and S.Wallon for useful discussions.


\eject
\appendix
\section{A useful formula involving generalized hypergeometric
functions ${}_3F_{2}$.}

In this appendix, one finds an appropriate summation for the hypergeometric function
\begin{equation}
{}_{3}F_{2}\!\left(\matrix {1,b+n,c-n}\\ {1+b+p, 1+c+q}\endmatrix ;
  {\displaystyle 1}\right)\nonumber
\end{equation}
for any integer $n$ and nonnegative integer values of $p$ and $q$.
Note that we can stick to nonnegative $n$ and obtain $-n$ by
interchanging $b$ and $c$ and $p$ and $q$ respectively.
The results are given in eq.(\ref{eq:resa1},\ref{eq:resa2}) or eq.(\ref{eq:res}).

The existence of the ${}_3F_{2}$
function is ensured by the strict positivity of the real part of the quantity
\begin{equation}
s\equiv(1+\!b\!+\!p)\!+\!(1\!+\!c\!+\!q)\!-\!1\!-\!(b\!+\!n)\!-
  \!(c\!-\!n)=1\!+\!p\!+\!q\nonumber\ .
\end{equation}
For our purpose, the
relevant values for the parameters of the ${}_3F_2$-function are any nonnegative
integer $n$ and $p,q\in\{0,1,2\}$.
An elementary method consists in expressing the function as a series
of quotients of $\Gamma$ functions, which reduces to a series of
rational fractions due to the particular values of its arguments. 
The latter are decomposed as a series of terms with
minimal denominators (which are first order polynomials in the
summation index). One can resum the series and one finally obtains a
finite sum whose terms are expressible as ratios of $\Gamma$ functions and possibly 
(depending on the relative values of $p$ and $n$) $\Psi$ functions. 
However, the number of contributing terms apparently grows as $n$, and one cannot
easily get by this method a simple expression for general $n$ and
small $p$ and $q$. This remark can be illustrated by applying a
transformation formula
(see ref.\cite{prudnikov}, formula (1) page 533):
\begin{equation}
{}_3F_2\!\left(\matrix {1,b+n,c-n}\\ {1+b+p, 1+c+q}\endmatrix ;
  {\displaystyle 1}\right)=\frac{\G(1\!+\!p\!+\!q)}{\G(1\!+\!n\!+\!q)}
\frac{\G(1\!+\!c\!+\!q)}{\G(1\!+\!c\!+\!p\!+\!q\!-\!n)}
{}_3F_2\!\left(\matrix {p\!-\!n,b\!+\!p,c\!-\!n}\\ 
{1\!+\!c\!+\!p\!+\!q\!-\!n, 1\!+\!b\!+\!p}\endmatrix ;
  {\displaystyle 1}\right)\ .
\end{equation}
One sees that in the case $n<p$, the r.h.s. of the preceeding equation
is an hypergeometric polynomial which has $n-p$ terms.
Thus a more sophisticated method has
to be developped.

Let us distinguish the cases {\bf (i)} $n>p$ and {\bf (ii)}
$n\leq p$.\\
\noindent
{\bf (i)} $n>p$.\\
This case can be obtained by immediate application of formula (6)
({\it Ibidem}, page 534). The condition of applicability is $\real(c+q)>0$ which
is satisfied in the case of interest in the core of the paper.
The result reads:
\begin{multline}
{}_{3}F_{2}\!\left(\matrix {1,b\!+\!n,c\!-\!n}\\ 
{1\!+\!b\!+\!p, 1\!+\!c\!+\!q}\endmatrix ;
  {\displaystyle 1}\right)=
\frac{\G(1\!+\!b\!+\!p)}{\G(b\!+\!n)}
\frac{\G(1\!+\!c\!+\!q)}{\G(c\!-\!n)}
\frac{\G(c\!-\!b\!-\!n\!-\!p)}{\G(1\!+\!c\!-\!b\!+\!q\!-\!n)}
B(1\!+\!p\!+\!q,n\!-\!p)\\
-\frac{(b+p)(c+q)}{(n-p)(c-b-n-p)}\,
{}_{3}F_{2}\!\left(\matrix {-p\!-\!q,1\!-\!b\!-\!p,1}\\ 
{1\!+\!n\!-\!p, 1\!+\!c\!-\!b\!-\!n\!-\!p}\endmatrix ;
  {\displaystyle 1}\right)
\label{eq:resa1}
\end{multline}
Note that the ${}_3F_2$ in the r.h.s. is a finite sum containing
$p\!+\!q$ terms, thanks to the parameter $-\!p\!-\!q$.

\noindent
{\bf (ii)} $n\leq p$.\\
This case is a bit more tricky. One uses formula (3) 
(see ref.\cite{prudnikov}), but one first needs to regularize by introducing
a small parameter $\epsilon$ such that:
\begin{equation}
{}_{3}F_{2}\!\left(\matrix {1,b+n,c-n}\\ {1+b+p, 1+c+q}\endmatrix ;
  {\displaystyle 1}\right)=
\lim_{\epsilon\rightarrow 0} {}_{3}F_{2}\!\left(\matrix {1+\epsilon,b+n,c-n}\\
    {1+b+p, 1+c+q}\endmatrix ;
  {\displaystyle 1}\right)\ .\nonumber
\end{equation}
This leads to:
\begin{multline}
{}_{3}F_{2}\!\left(\matrix {1+\epsilon,b+n,c-n}\\
    {1+b+p, 1+c+q}\endmatrix ;
  {\displaystyle 1}\right)=
\G(1\!+\!b\!+\!p)\G(1\!+\!c\!+\!q)\G(-\epsilon)
\Bigg\lbrace\frac{\G(c\!-\!b\!-\!2n)}
{\G(1\!+\!p\!-\!n)\G(1\!+\!c\!-\!b\!+\!q\!-\!n)
\G(b\!+\!n\!-\!\epsilon)}\times\\
\times\sum_{k=0}^{\infty}
\frac{(b\!+\!n)_k (b\!-\!c\!+\!n\!-\!q)_k (n\!-\!p)_k}
{\G(1\!+\!k)(b\!+\!n\!-\!\epsilon)_k (1\!+\!b\!-\!c\!+\!2n)_k}
+\left[\matrix {b\leftrightarrow c}\\
{n\leftrightarrow -n}\\ {p\leftrightarrow q}
\endmatrix
\right]
\Bigg\rbrace\ ,
\end{multline}
where we have employed the classical notation $({\mathcal
A})_k=\G({\mathcal A}\!+\!k)/\G({\mathcal A})$, and the square
brackets is a short notation for the term obtained by the indicated
exchange of parameters.
The next step is to take the limit $\epsilon\rightarrow 0$.
One uses the fact that $\G({\mathcal A}\!+\!\epsilon)=
\G({\mathcal A})(1\!+\!\epsilon\,\Psi({\mathcal A}))\!+\!{\mathrm o}(\epsilon)$
to obtain:
\begin{multline}
{}_{3}F_{2}\!\left(\matrix {1+\epsilon,b+n,c-n}\\
    {1+b+p, 1+c+q}\endmatrix ;
  {\displaystyle 1}\right)=
-\frac{\G(1\!+\!b\!+\!p)}{\G(b\!+\!n)}\frac{\G(1\!+\!c\!+\!q)}{\G(c\!-\!n)}
\Bigg\lbrace\frac{\G(c\!-\!b\!-\!2n)}{\G(1\!+\!p\!-\!n)
\G(1\!+\!c\!-\!b\!+\!q\!-\!n)}\times\\
\times\sum_{k=0}^{p-n}\frac{(b\!-\!c\!-\!q\!+\!n)_k(n\!-\!p)_k}
{\G(1\!+\!k)(1\!+\!b\!-\!c\!+\!2n)_k}
\left(\Psi(b\!+\!n\!+\!k)\!-\!\Psi(b\!+\!n)\right)\\
+\frac{\G(c\!-\!b\!-\!2n)}{\G(1\!+\!p\!-\!n)
\G(1\!+\!c\!-\!b\!+\!q\!-\!n)}
\frac{\G(1\!+\!p\!+\!q)}{\G(1\!+\!n\!+\!q)}
\frac{\G(1\!+\!b\!-\!c\!+\!2n)}{\G(1\!+\!b\!-\!c\!+\!n\!+\!p)}
\Psi(b\!+\!n)
+\left[\matrix {b\leftrightarrow c}\\
{n\leftrightarrow -n}\\ {p\leftrightarrow q}
\endmatrix
\right]\Bigg\rbrace\ .
\end{multline}
We wrote it in this form in order to isolate the special functions
by using the summation formula (\ref{eq:appb}) in appendix {\bf B}.
After a bunch of straightforward manipulations, 
we are led to the following result:
\begin{multline}
{}_{3}F_{2}\!\left(\matrix {1+\epsilon,b+n,c-n}\\
    {1+b+p, 1+c+q}\endmatrix ;
  {\displaystyle 1}\right)=
-\frac{\G(1\!+\!b\!+\!p)}{\G(b\!+\!n)}
\frac{\G(1\!+\!c\!+\!q)}{\G(c\!-\!n)}
\Bigg\lbrace
(-1)^{p-n}\left(\matrix{p\!+\!q}\\{p\!-\!n}\endmatrix\right)\times\\
\times\frac{\G(c\!-\!b\!-\!n\!-\!p)}{\G(1\!+\!c\!-\!b\!+\!q\!-\!n)}
\left(\Psi(b\!+\!n)\!-\!\Psi(c\!-\!n)\right)
+\Bigg\lbrack\frac{\pi}{\sin\pi b}
\frac{\G(c\!-\!b\!-\!p\!-\!n)\G(b\!+\!n)}{\G(1\!+\!q\!+\!n)}\times\\
\times\sum_{k=0}^{p\!-\!n\!-\!1}\frac{1}{n\!-\!p\!+\!k}
\frac{1}{\G(1\!+\!k)}
\frac{\G(1\!+\!q\!+\!n\!+\!k)\G(1\!-\!b\!-\!p\!+\!k)}
{\G(1\!+\!c\!-\!b\!+\!q\!-\!p\!+\!k)}
+\left[\matrix {b\leftrightarrow c}\\
{n\leftrightarrow -n}\\ {p\leftrightarrow q}
\endmatrix
\right]
\Bigg\rbrack\Bigg\rbrace\ .
\label{eq:resa2}
\end{multline}
The result has been put in a form which shows that
if $c=1-b$, as it is the case in the core of this paper, the difference of
$\Psi$ reduces to a tangent, through the identity:
\begin{equation}
\Psi(1-x)-\Psi(x)=\frac{\pi}{\tan\pi x}\ .
\end{equation}

One of the most interesting points in these formulae is that 
the number of terms is at most equal to $p+q$, regardless the value of
$n$, making them particularly adapted for our purpose.

Note that in the two cases $n\!=\!0$ and $\{p,q\}\!=\!0,1$,
the obtained formula matches with formerly calculated expressions, see 
for instance ref.\cite{prudnikov}.\\

We can obtain a plethora of such formulae by using the relations
between the various ${}_3F_2$ functions. 
A particularly interesting one is the following:
\begin{multline}
{}_{3}F_{2}\!\left(\matrix {1,b+n,c-n}\\ {1+b+p, 1+c+q}\endmatrix ;
  {\displaystyle 1}\right)=
\frac{\G(1\!+\!b\!+\!p)\G(1\!+\!c\!+\!q)\G(c\!-\!b\!-\!n\!-\!p)}
  {\G(b\!+\!n)\G(c\!-\!n)\G(1\!+\!c\!-\!b\!-\!n\!+\!q)}
  \times\\
 \times\sum_{i=0}^{p\!+\!q}\frac{(-1)^i}{n\!-\!p\!+\!i}
\left(\matrix{p+q}\\i\endmatrix\right)
\left(1-\frac{\G(b\!+\!n\!+\!i)}
  {\G(b\!+\!p)}
\frac{\G(c\!-\!n)}{\G(c\!-\!p\!+\!i)}\right)\ .
\label{eq:res}
\end{multline}
(We do not reproduce the proof here, but for $p<n$, it mainly relies on the use
of formula (25) on page 108 in ref.\cite{luke}).
Note that the formula is trivially analytical for $n>p$, but also in the limit
$n\!\equiv\!m\!\leq\!p$: all the terms are analytical but for one pole
appearing as a denominator $1/(n-p+i)$ for the appropriate value of $i$. 
However, its residue vanishes and so the expression in the r.h.s is
finite. 
We checked numerically that the obtained expression
is correct for the values of $p,q$ of interest. 
The property is not trivial since the asymptotic 
behaviour of the given ${}_{3}F_{2}$ does not satisfy the hypotheses
for Carlson's theorem and hence there are infinitely many ways to continue
analytically $n>p$.


\section{A summation formula.}
\label{app:sommation}

For the needs of appendix {\bf A}, we simplify the following expression, 
getting rid of the $\Psi$-functions:
\begin{equation}
\sum_{k=0}^{l}\frac{1}{\Gamma(1\!+\!k)}\frac{(-l)_k(\alpha)_k}{(\gamma)_k}
\left\{\Psi(\beta\!+\!k)\!-\!\Psi(\beta)\right\}\nonumber\ .
\end{equation}
One notes that:
\begin{equation}
\Psi(\beta\!+\!k)\!-\!\Psi(\beta)=\sum_{i=0}^{k-1}\frac{1}{\beta\!+\!i}\ ,
\nonumber
\end{equation}
and one defines the function:
\begin{equation}
f(z)=\sum_{k=0}^{l}\frac{1}{\Gamma(1\!+\!k)}\frac{(-l)_k(\alpha)_k}{(\gamma)_k}
\left\{
\sum_{i=0}^{k-1}\frac{z^{\beta+i}}{\beta\!+\!i}\right\}\ ,
\end{equation}
$z$ being an arbitrary parameter. The quantity we need to compute is $f(1)$.
This goes as follows.
The derivative of $f$ can be summed and we obtain the difference of 
two hypergeometric ${}_2F_1$ functions:
\begin{equation}
f^\prime(z)=\frac{z^{\beta-1}}{z\!-\!1}\left\{
{}_2F_1(-l,\alpha,\gamma,z)\!-\!{}_2F_1(-l,\alpha,\gamma,1)\right\}\ .
\end{equation}
One uses a well-known transformation \cite{gradsteyn} for the first term in the
r.h.s. of the preceeding equation:
\begin{equation}
{}_2F_1(-l,\alpha,\gamma,z)=(1\!-\!z)^l\frac{\G(\gamma)\G(\alpha\!+\!l)}
{\G(\alpha)\G(\gamma\!+\!l)}
{}_2F_1(-l,\gamma\!-\!\alpha,1\!-\!l\!-\!\alpha,1/(1-z))\ .
\end{equation}
The integration over $z$ can then be performed safely for 
$\real(\beta)>0$.
After a few easy manipulations, the result can be written in the
following compact form:
\begin{equation}
\sum_{k=0}^{l}\frac{1}{\G(1\!+\!k)}\frac{(-l)_k(\alpha)_k}{(\gamma)_k}
\left\{\Psi(\beta\!+\!k)\!-\!\Psi(\beta)\right\}
=\frac{(\alpha)_l}{(\beta)_l(\gamma)_l}
\sum_{k=0}^{l-1}\frac{1}{k\!-\!l}\frac{\G(1\!+\!l)}{\G(1\!+\!k)}
\frac{(\gamma\!-\!\alpha)_k(1\!-\!\beta\!-\!l)_k}{(1\!-\!\alpha\!-\!l)_k}
\ .
\label{eq:appb}
\end{equation}
Note the following particular case occuring when $\alpha\equiv\gamma$:
\begin{equation}
\sum_{k=0}^l (-1)^{k+1}\left(\matrix{l}\\k\endmatrix\right)
\Psi(\beta+k)=B(\beta,l)\ .
\end{equation}


\section{Some particular cases.}

In this appendix, we compute the difference of ${}_3G_2$-functions in
eq.(\ref{eq:diffG}), using formulae (\ref{eq:resa1},\ref{eq:resa2})
or alternatively formula (\ref{eq:res}). 
To fix the notation:
\begin{equation}
{\mathcal G}_{pq}(n)\equiv
{}_{3}G_{2}\!\left(\matrix {1,\gamma\!+\!\alpha\!-\!1\!+\!n,\gamma\!-\!\alpha\!-\!n}\\ 
{\gamma\!+\!\alpha\!+\!p, \gamma\!-\!\alpha\!+\!1\!+\!q} \endmatrix ;
  {\displaystyle 1}\right)
-{}_{3}G_{2}\!\left(\matrix {1,\!\alpha\!+\!n,1\!-\!\alpha\!-\!n}\\ 
{1\!+\!\alpha\!+\!p, 2\!-\!\alpha\!+\!q} \endmatrix ;
  {\displaystyle 1}\right)\ .
\end{equation}
We recall that ${}_3G_2$ is related to ${}_3F_2$ through the relation:
\begin{equation}
{}_{3}G_{2}\!\left(\matrix {a_1,a_2,a_3}\\
{b_1,b_2} \endmatrix ; {\displaystyle z}\right)
\equiv
\frac{\G(a_1)\G(a_2)\G(a_3)}{\G(b_1)\G(b_2)}\;
{}_{3}F_{2}\!\left(\matrix {a_1,a_2,a_3}\\ 
{b_1,b_2} \endmatrix ; {\displaystyle z}\right)\ .
\end{equation}
A few of the following expressions can be obtained using ref.\cite{prudnikov}.
For the others, we use the quoted formulae, 
with the parameter values $a\equiv\gamma\!+\!\alpha\!-\!1$ and 
$b\equiv\gamma\!-\!\alpha$ (resp. $a\equiv\alpha$ and $b\equiv 1\!-\!\alpha$).
We only display the result for $n=0$ and $n=2$ although a general
expression can be written.


\begin{equation}
{\mathcal G}_{00}(n=0)={\frac{1}{2\alpha\!-\!1}\left\{\pi \,\cot\pi\alpha\!+\! 
\Psi(\gamma\!+\!\alpha\!-\!1)\!-\!\Psi(\gamma\!-\!\alpha)\right\}}
\nonumber
\end{equation}

\begin{equation}
{\mathcal G}_{00}(n=2)=
\frac{\gamma\!-\!1}{(\gamma\!-\!\alpha\!-\!1)(\gamma\!-\!\alpha\!-\!2)}
\nonumber
\end{equation}


\begin{equation}
{\mathcal G}_{01}(n=0)={\frac{(\gamma\!-\!1)}{2\,{{(\alpha\!-\!1)
}^2}(\gamma\!-\!\alpha)}
\!-\!\frac{\pi 
\cot\pi\alpha\!+\!\Psi(\gamma\!+\!\alpha\!-\!1)\!-
\!\Psi(\gamma\!-\!\alpha) }
{2\,{{(\alpha\!-\!1) }}
(2\alpha\!-\!1) }}
\nonumber
\end{equation}

\begin{equation}
{\mathcal G}_{01}(n=2)=
\frac{(\gamma\!-\!1)(6\alpha^2\!-\!3\alpha\gamma\!+\!\gamma(\gamma\!-\!2))}
{6\alpha(1\!-\!\alpha)(\alpha\!-\!\gamma)(1\!+\!\alpha\!-\!\gamma)
(2\!+\!\alpha\!-\!\gamma)}
\nonumber
\end{equation}

\begin{equation}
{\mathcal G}_{10}(n=0)=
\frac{\gamma\!-\!1}{2\alpha^2(\gamma\!+\!\alpha\!-1)}
\!+\!\frac{
\pi\cot\pi\alpha\!+\!\Psi(\gamma\!+\!\alpha\!-\!1)\!-\!\Psi(\gamma\!-\!\alpha)}
{2\alpha(2\alpha\!-\!1)}
\nonumber
\end{equation}

\begin{equation}
{\mathcal G}_{10}(n=2)=
\frac12\left\{
\frac{1}{(\gamma\!-\!\alpha\!-\!1)(\gamma\!-\!\alpha\!-\!2)}
-\frac{1}{\alpha(1\!+\!\alpha)}
\right\}
\nonumber
\end{equation}


\begin{equation}
{\mathcal G}_{11}(n=0)=\frac{\frac{1}{\alpha\!-\!1}\!+\!\frac{1}{\alpha}\!+\!
\frac{\alpha}{\gamma\!-\!\alpha}\!+\!\frac{\alpha\!-\!1}{\gamma\!+\!\alpha\!-\!1}
\!-\!\pi\cot\pi\alpha\!+\!\Psi(\gamma\!-\!\alpha)\!-\!\Psi(\gamma\!+\!\alpha)}
{2\alpha(\alpha\!-\!1)(2\alpha\!-\!1)}
\nonumber
\end{equation}

\begin{equation}
{\mathcal G}_{11}(n=2)=-\frac16\left(
\frac{2}{\alpha(1\!-\!\alpha^2)}+\frac{1}{\alpha\!-\!\gamma}-
\frac{2}{1\!+\!\alpha\!-\!\gamma}
+\frac{1}{2\!+\!\alpha\!-\!\gamma}\right)
\nonumber
\end{equation}

\begin{multline}
{\mathcal G}_{12}(n=0)=
-\frac{1}{8\alpha(1\!-\!\alpha)(2\alpha\!-\!1)(2\alpha\!-\!3)}
\Bigg\lbrace 4\!-\!\frac{6}{\alpha}\!-\!\frac{(1\!+\!\alpha)(2\!+\!\alpha)}
{(1\!-\!\alpha)(2\!-\!\alpha)}
+\frac{12\alpha(\gamma\!-\!1)}{(1\!-\!\alpha)(\gamma\!-\!\alpha)}+\\
+2\frac{3\!-\!2\alpha}{\gamma\!+\!\alpha\!-\!1}+
\frac{(\gamma\!+\!\alpha)(\gamma\!+\!\alpha\!+\!1)}{(\gamma\!-\!\alpha)(\gamma\!-\!\alpha\!+\!1)}
\Bigg\rbrace
+\frac34\frac{\pi\cot\pi \alpha\!+\!\Psi(\gamma\!+\!\alpha\!-\!1)\!-\!\Psi(\gamma\!-\!\alpha)}
{\alpha(1\!-\!\alpha)(2\alpha\!-\!1)(2\alpha\!-\!3)}
\nonumber
\end{multline}

\begin{equation}
{\mathcal G}_{12}(n=2)=
-\frac{1}{24} 
\left\{\frac{6}{(\alpha\!-\!2)(\alpha\!-\!1)\alpha(\alpha\!+\!1)}\!+\! 
\frac{3}{\alpha\!-\!\gamma}\!+\!\frac{1}{2\!+\!\alpha\!-\!\gamma}\!+\!
\frac{3}{-1\!-\!\alpha\!+\!\gamma}\!+\!\frac{1}{1\!-\!\alpha\!+\!\gamma}\right\}
\nonumber
\end{equation}

\begin{multline}
{\mathcal G}_{21}(n=0)=
-\frac{1}{8(-1\!+\!\alpha)\alpha
     (-1\!+\!2\alpha)(1\!+\!2\alpha)}
\Bigg\lbrace
1\!-\!{\frac{6}{\alpha}}\!-\!{\frac{6}{1\!+\!\alpha}}\!-\!
{\frac{2(2\!+\!\alpha)}{-1\!+\!\alpha}}\!+\! 
     {\frac{6(1\!+\!2\alpha)(-1\!+\!\gamma)}{(1\!+\!\alpha)(\alpha\!+\!\gamma)}}\\
-\frac{(\alpha\!-\!1)(\alpha\!-\!8)\!+\!\gamma(\gamma\!-\!2\alpha\!-\!15)} 
{(-1\!+\!\alpha\!+\!\gamma)(\alpha\!+\!\gamma)}
+{\frac{2(1\!+\!\alpha\!+\!\gamma) }{\alpha\!-\!\gamma}}
\Bigg\rbrace
- 
\frac34\frac{\pi\cot\pi \alpha\!-\!\Psi(-\alpha\!+\!\gamma) \!+\! 
\Psi(-1\!+\!\alpha\!+\!\gamma)}
{(-1\!+\!\alpha)\alpha
 (-1\!+\!2\alpha)(1\!+\!2\alpha)}
\nonumber
\end{multline}

\begin{multline}
{\mathcal G}_{21}(n=2)=
-\frac{
    (-1\!+\!\gamma) 
   (3\alpha^2(1\!+\!\alpha)(5\!+\!7\alpha)\!+\!2\gamma \!-\! 
    \alpha(6\!+\!\alpha(29\!+\!27\alpha))\gamma + 
   (1\!+\!2\alpha)(-1\!+\!5\alpha)\gamma^2}
{12(-1\!+\!\alpha)\alpha^2(1\!+\!\alpha)^2(1\!+\!2\alpha)
         (\alpha\!-\!\gamma)(1\!+\!\alpha\!-\!\gamma)(2\!+\!\alpha\!-\!\gamma)} + \\
-\frac{
  \pi\cot\pi\alpha- 
           \Psi(-2\!-\!\alpha\!+\!\gamma)\!+\!\Psi(1\!+\!\alpha\!+\!\gamma)}
  {4\alpha(1\!+\!\alpha)(1\!+\!2\alpha)
   (3\!+\!2\alpha)}
\nonumber
\end{multline}





\section{Comparison with the calculation of Evanson and Forshaw.}

We write here another expression for eq.(\ref{eq:conforme}) in order
to try to match at this level our calculation with the one of 
ref.\cite{Evanson:1999zb}.
We perform the following transformation on the (antiholomorphic) 
${}_2G_1$-functions in formula (\ref{eq:conforme}) \cite{gradsteyn}:
\begin{multline}
{}_2G_1(\alpha,\beta,\gamma,z)=\frac{\pi}{\sin\pi(\gamma\!-\!\alpha\!-\!\beta)}
\Bigg\lbrace
\frac{1}{\G(\gamma\!-\!\alpha)\G(\gamma\!-\!\beta)}
{}_2G_1(\alpha,\beta,1\!+\!\alpha\!+\!\beta\!-\!\gamma,1\!-\!z)\\
-\frac{z^{1\!-\!\gamma}(1\!-\!z)^{\gamma\!-\!\alpha\!-\!\beta}}
{\G(1\!-\!\alpha)\G(1\!-\!\beta)}
{}_2G_1(1\!-\!\alpha,1\!-\!\beta,1\!-\!\alpha\!-\!\beta\!+\!\gamma,1\!-\!z)
\Bigg\rbrace\ .
\end{multline}
One makes use of the relations:
\begin{align}
a_0\!+\!a_1\!+\!1&=2b_1\nonumber\\
\tilde a_0\!+\!\tilde a_1\!+\!1&=2\tilde b_1\ .
\end{align}
The result reads:
\begin{multline}
-2i\pi\frac{\mu}{\sin\pi b_1}\frac{\sin\pi(b_1\!-\!a_0)}{\sin\pi b_1}
\Bigg\lbrace
\alpha^{1\!-\!b_1}\bar{\alpha}^{1\!-\!\tilde{b}_1}
\frac{\sin\pi(\tilde{b}_1\!-\!\tilde{a}_0)}{\G(1\!-\!a_0)\G(1\!-\!a_1)}\;
{}_2G_1(b_1\!-\!a_0,b_1\!-\!a_1,b_1,1\!-\!\alpha)
{}_2G_1(\tilde b_1\!-\!\tilde
a_0,\tilde b_1\!-\!\tilde a_1,2\!-\!\tilde b_1,\bar\alpha)\\
+(1\!-\!\alpha)^{1-b_1}(1\!-\!\bar\alpha)^{1\!-\!\tilde b_1}
\frac{\sin\pi(b_1\!-\!a_0)}{\G(1\!-\!\tilde a_0)\G(1\!-\!\tilde a_1)}\;
{}_2G_1(b_1\!-\!a_0,b_1\!-\!a_1,2\!-\!b_1,1\!-\!\alpha)
{}_2G_1(\tilde b_1\!-\!\tilde
a_0,\tilde b_1\!-\!\tilde a_1,\tilde b_1,\bar\alpha)\\
-\alpha^{1-b_1}(1\!-\!\bar\alpha)^{1-\tilde b_1}
\frac{1}{\pi}\sin^2\pi(\tilde b_1\!-\!\tilde a_0)
\frac{\G(\tilde a_0)\G(\tilde a_1)}{\G(1\!-\!a_0)\G(1\!-\!a_1)}\;
{}_2G_1(b_1\!-\!a_0,b_1\!-\!a_1,b_1,1\!-\!\alpha)
{}_2G_1(\tilde b_1\!-\!\tilde
a_0,\tilde b_1\!-\!\tilde a_1,\tilde b_1,\bar\alpha)\\
-(1\!-\!\alpha)^{1-b_1}\bar\alpha^{1-\tilde b_1}\frac{1}{\pi}\sin\pi(b_1\!-\!a_0)
\sin\pi(\tilde b_1\!-\!\tilde a_0)\;
{}_2G_1(b_1\!-\!a_0,b_1\!-\!a_1,2\!-\!b_1,1\!-\!\alpha)
{}_2G_1(\tilde b_1\!-\!\tilde
a_0,\tilde b_1\!-\!\tilde a_1,2\!-\!\tilde b_1,\bar\alpha)
\Bigg\rbrace\ .
\end{multline}
In our particular case, $\alpha$ is real and so $\bar\alpha=\alpha$.

Furthermore, if one is only interested by the case $n=0$ and, say, the non-flip
helicity amplitude, then $a_0=\tilde a_0$, $a_1=\tilde a_1$ and
$b_1=\tilde b_1$. One will have to integrate this expression over
$\alpha$ after having multiplied it by a symmetric fonction under the
exchange $\alpha\rightarrow 1-\alpha$.

Hence one sees on the previous formula that the first two terms 
give similar contributions after integration and are proportional to
the one quoted in ref.\cite{Evanson:1999zb}, eq.(3.46).
The two other terms do not appear in the latter. 
So at this stage the intermediate forms differ, but the
integrated results are identical. However, we failed to explain why.


%


\newpage

\noindent
{\bf\Large Figure captions.}\\
\vskip 2cm
\noindent
{\bf Fig.1} Scattering process.\\

\noindent
{\it $1$ and $2$ label the colliding
objects. The wavy lines represent exchanged reggeized gluons which
interact through the BFKL kernel. The arrows give the momentum flux.}

\vskip 0.3cm
\noindent
\hrulefill
\vskip 0.7cm
\noindent
{\bf Fig.2} Diagram contributing to the impact factor $\phi_1$.\\

\noindent
{\it The dashed lines represent (virtual or real) photons which
resolve into a quark-antiquark pair. The wavy
lines stand for off-shell gluons. Four such diagrams (corresponding to
each possible insertion of the two exchanged gluons) have to be
taken into account although (see text) only two of them 
effectively contribute to the amplitude.}

\newpage

\begin{center}
\vskip 1cm
\epsfig{file=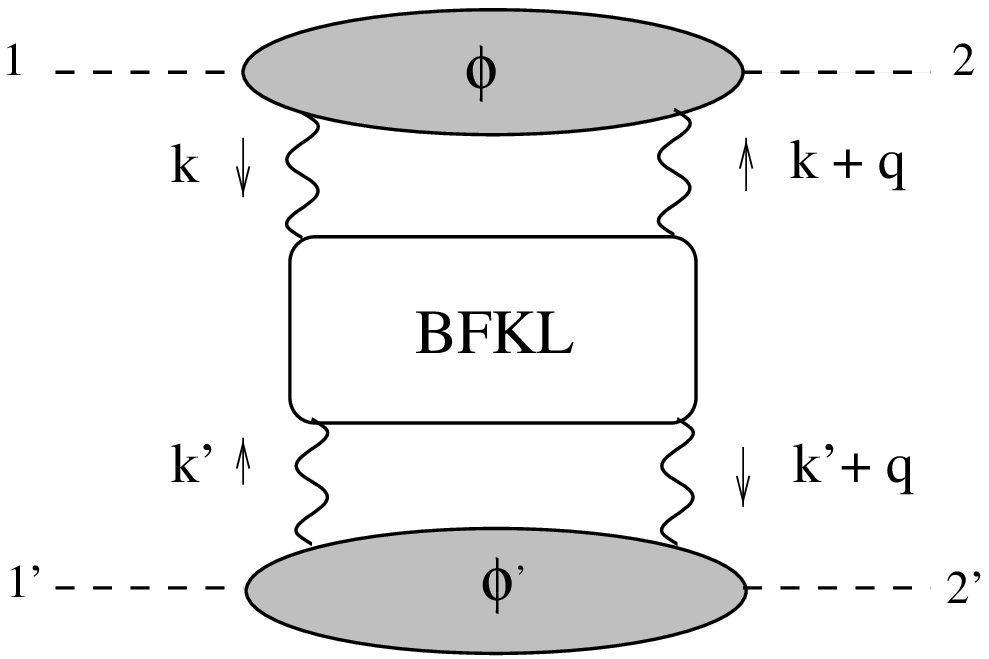,width=12cm}
\vskip 1cm
{\bf\Large Fig.1}
\vskip 2cm
\noindent
\hrulefill
\vskip 2cm
\epsfig{file=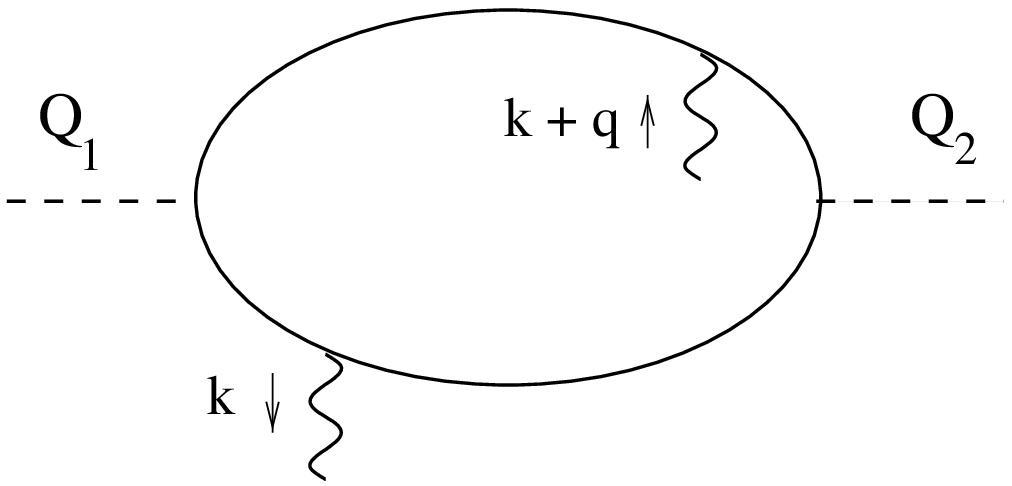,width=12cm}
\vskip 1cm
{\bf\Large Fig.2}
\end{center}

\end{document}